# Effect of Anodizing Parameters on Corrosion Resistance of Coated Purified Magnesium

Sohrab Khalifeh[a,1], and T. David Burleigh[2]

[a] Department of Materials and Metallurgical Engineering Department, New Mexico Institute of Mining and Technology, Socorro, New Mexico, USA

**Abstract**

Magnesium and its alloys are being considered for a biodegradable biomaterials. However, high and uncontrollable corrosion rates have limited the use of magnesium and its alloys in biological environments. In this research, high purified magnesium (HP-Mg) was coated with stearic acid in order to improve the corrosion resistance of magnesium. Anodization and immersion in stearic acid were used to form a hydrophobic layer on magnesium substrate. Different DC voltages, times, electrolytes, and temperatures were tested. Electrochemical impedance spectroscopy and potentiodynamic polarization were used to measure the corrosion rates of the coated HP-Mg. The results showed that optimum corrosion resistance occurred for specimens anodized at +4 volts for 4 minutes at 70°C in borate benzoate. The corrosion resistance was temporarily enhanced by 1000x.

## 1. Introduction

Biodegradable materials are being considered for temporary implants to cut the cost and the problems associated with the retrieval surgery [1, 2]. Magnesium and its alloys are ideal for orthopedic implants due to their biocompatibility, low stress shielding, and good strength to weight ratio. Also, magnesium is gradually dissolved in the human body and is removed [3-6]. However, using magnesium and its alloys as biodegradable implants have been limited due to their rapid corrosion rate in the biological environment. Magnesium is dissolved via an electrochemical reaction in the presence of water due to the thermodynamic instability of magnesium. This chemical reaction produces magnesium hydroxide and hydrogen gas [7]. The high concentration of chloride ions in the biological environment can lead to the pitting corrosion of magnesium, and consequently can cause serious problems such as the loss of mechanical properties and the disappearance of the implant during the healing period along with hydrogen bubbles and caustic burning [2] [8-12].

---

[1] Corresponding author. *Email address:* skhalife@nmt.edu
[2] burleigh@nmt.edu

Many investigations have been conducted in order to control and slow the corrosion rate of magnesium and magnesium alloys. Coating is one of the methods that has been considered to achieve this goal [13-18]. However, the coating layer should also be dissolvable and biocompatible.

In this research, biocompatible and non-toxic stearic acid (SA) ($CH_3(CH_2)_{16}COOH$) [19] was selected in order to control the corrosion rate of magnesium. The magnesium substrate must be treated to ensure sufficient adhesion to the SA layer. Ng et al. [20] conducted hydrothermal treatments to form an $Mg(OH)_2$ layer which provided sites to attach the SA coating. In this research, anodization with a DC voltage and immersion in SA were used to form a hydrophobic layer. The effect of different anodization parameters on the corrosion behavior of magnesium is reported herein.

## 2. Material and Methods

### 2.1. Sample Preparation

High purified magnesium (HP-Mg) rods (see Table 1) were cut into 10 mm long cylinders (D=18 mm) and cold mounted in epoxy. The specimens were polished with sandpaper (1200 grit) and oil-based diamond slurry (1 $\mu$m), degreased with ethanol, washed with deionized water, and dried with flushing compressed air.

*Table 1. Composition for HP-Mg used in this work (weight percent).*

| Material | Mg | Al | Zn | Ca | Si | Mn | Fe | Ni | Cu | Zr | Na |
|---|---|---|---|---|---|---|---|---|---|---|---|
| HP-Mg | 99.97 | 0.002 | 0.005 | 0.001 | 0.014 | 0.001 | 0.003 | 0.002 | >0.0002 | >0.002 | 0.002 |

The coating process was conducted on the magnesium specimens with a two-step process. First the specimens were anodized to form a $Mg(OH)_2$/MgO layer. The anodization processes were performed at 30, 50, 70°C for 2-4 minutes with 4-6 volts (DC) in the aqueous electrolytes (50% NaOH, 1 M KOH, and borate benzoate (NaBz)) using a TecNu power supply (Dca: 25/12-1Z) to provide multiple ranges of anodizing. HP-Mg was connected to a positive terminal of the power supply and the platinum wire as a counter electrode was connected to the negative terminal. The borate benzoate electrolyte was composed of 60 g/L NaOH, 25 g/L $Na_2B_4O_7$, 20 g/L $H_3BO_3$, 3 g/L NaBz (sodium benzoate: $C_6H_5COONa$) [21] and heated at 30°C on a hot plate.

The second step was immersion (1-60 min) in 0.05-0.15 mol/L SA in ethanol at room temperature. After immersion, the specimens were dried in the fume hood for 24 hours.

## 2.2. Evaluation of Hydrophobic Layer

Studies of the surface morphology of the anodized specimens were carried out using scanning electron microscopy (SEM) (Hitachi S-3200). The water contact angles were measured at room temperature using the Kruss Drop Shape Analyzer (DSA25) in order to evaluate the hydrophobicity of coated layer.

## 2.3. Corrosion Tests

The corrosion tests on the polished bare and coated HP-Mg specimens were conducted in simulated body fluid (SBF) at 37°C. The SBF was composed according to the protocol for preparing SBF as describe by Kokubo [22], and the concentrations are shown in Table 2.

*Table 2. Ion concentration of SBF [22].*

|  | Simulated Body Fluid (SBF)(mM) |
|---|---|
| Na+ | 142.0 |
| K+ | 5.0 |
| Mg+ | 1.5 |
| Ca+ | 2.5 |
| Cl- | 148.8 |
| HCO$_3$- | 4.2 |
| HPO$_4$$^{2-}$ | 1.0 |
| SO$_4$$^{-2}$ | 0.5 |

Electrochemical impedance spectroscopy (EIS) and potentiodynamic polarization (PDP) were measured in a cell containing 275 ml of SBF at 37°C and the specimen area exposed was 2.55 cm². For all electrochemical measurements, the three-electrode method was used with a PARSTAT2263 potentiostat. The KCl saturated calomel electrode (SCE) and platinum wire were used as a reference and counter electrode respectively. Three specimens were used for each condition to confirm the reproducibility of the EIS and PDP measurements. EIS test was performed after 5-10 minutes immersion in SBF and run in the frequency range from 100 kHz to 100 mHz. The PDP test was run after the EIS test. The initial potential was scanned at a rate of 10 mV/sec from -250 mV relative to open circuit potential (OCP) and stopped at +1600 mV OCP. The corrosion current density, $i_{corr}$ $(A/cm^2)$ at $E_{corr}$ was obtained by extrapolating the cathodic and anodic Tafel regions. EIS tests were performed to study the long-term corrosion behavior of the coated HP-Mg for up to 3 days. The corrosion current density, $i_{corr}$ $(A/cm^2)$ from each PDP test was converted to the corrosion rate (mm/year) by the following conversion equation;

$$C.R._{EIS/PDP} = i_{corr}\left(\frac{A}{cm^2}\right)\left(\frac{C}{A\cdot sec}\right)\left(\frac{e^{-1}}{1.602\times 10^{-19}C}\right)\left(\frac{Mg\ atom}{2e^-}\right)\left(\frac{mol\ Mg}{6.022\times 10^{23}\ Mg\ atom}\right)$$

$$\left(\frac{24.305g\ Mg}{mol\ Mg}\right)\left(\frac{cm^3}{1.74g}\right)\left(\frac{10mm}{cm}\right)\left(\frac{3600sec}{h}\right)\left(\frac{24h}{day}\right)\left(\frac{365day}{year}\right) \quad (1)$$

$$= 22.83\times 10^3\ i_{corr}\ mm\cdot year^{-1}$$

## 3. Results and Discussions

### 3.1. Characterization of the Coated Layer

The HP-Mg specimens were covered with a gray layer of Mg(OH)$_2$/MgO after anodization process at +4, +5 and +6 Volts (DC) for 2, 3 and 4 minutes. The surface morphology of anodized HP-Mg is illustrated in Figure 1 at different magnifications. Figure 1 b-d show that the anodization processes provided the flaky morphology on the magnesium substrate. The flaky structure on the surface allows SA to penetrate the HP-Mg substrate, also increasing the surface area of the magnesium binding with SA.

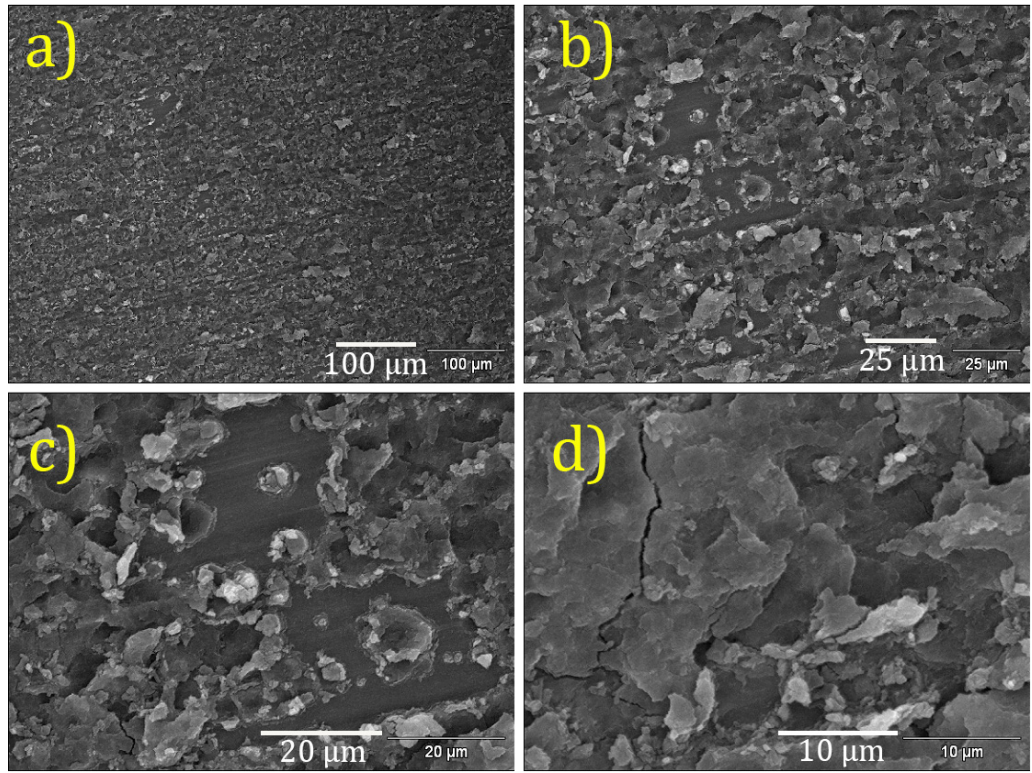

Figure 1. SEM images of anodized HP-Mg at different magnifications (+4 Volts for 3 min).

The hydrophobicity of the polished bare and coated HP-Mg were investigated. The highest hydrophobicity was obtained with the coated magnesium, which was polished (1200 grit), anodized at +5 volts for 3 minutes in borate benzoate at 30°C, and then immersed in 0.05 mol/L SA in ethanol for 60 minutes (Figure 2). The water contact angle of the polished bare and the coated HP-Mg were 54° and 164° respectively. The water contact angles of all the coated HP-Mg at different DC voltages (4-6 volts), times (2-4 min) of anodization at 30°C are shown in Table 3.

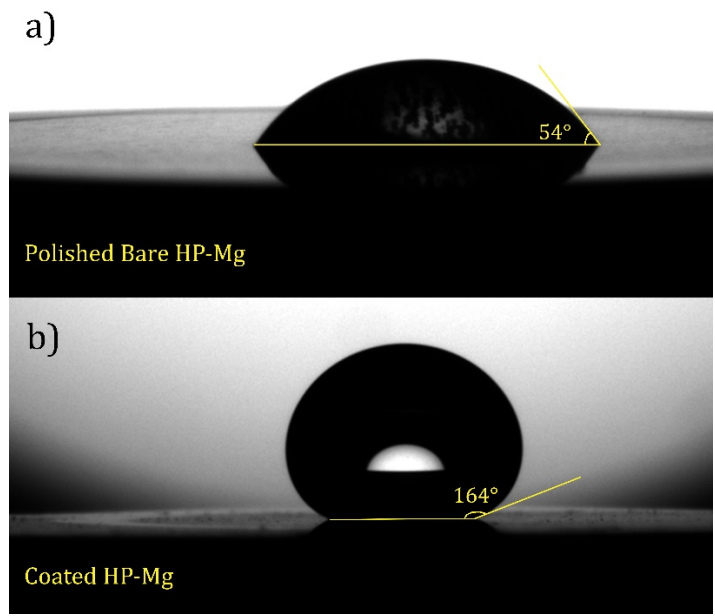

Figure 2. The water contact angle of a water drop, a) Polished bare HP-Mg (1200 grit), b) Coated HP-Mg.

### 3.2. Corrosion Behavior (EIS and PDP Results)

In order to investigate the influence of voltage, time, and temperature, the polished (1200 grit) magnesium specimens were anodized with +4, +5, and +6 volts for 2, 3, and 4 minutes at 30, 50, and 70°C in borate benzoate, then immersed in 0.05 mol/L SA in ethanol for 60 minutes. The PDP and EIS measurements were performed in SBF at 37°C. The PDP curves illustrate a different anodic polarization behaviors for different specimens in SBF. The corrosion density, $i_{corr}$ $(A/cm^2)$ was extrapolated from the PDP curves, then converted to the corrosion rate (mm/year) with equation 1. The results are shown in Figure 3.

Table 3. The water contact angles of the coated HP-Mg specimens at different DC voltages and times (30°C).

| Anodization Voltage | Time of Anodization (min) | Contact Angles |
|---|---|---|
| - | - | 54° [3] |
| +4 Volts | 2 | 113° |
| | 3 | 145° |
| | 4 | 134° |
| +5 Volts | 2 | 148° |
| | 3 | 164° |
| | 4 | 151° |
| +6 Volts | 2 | 140° |
| | 3 | 132° |
| | 4 | 135° |

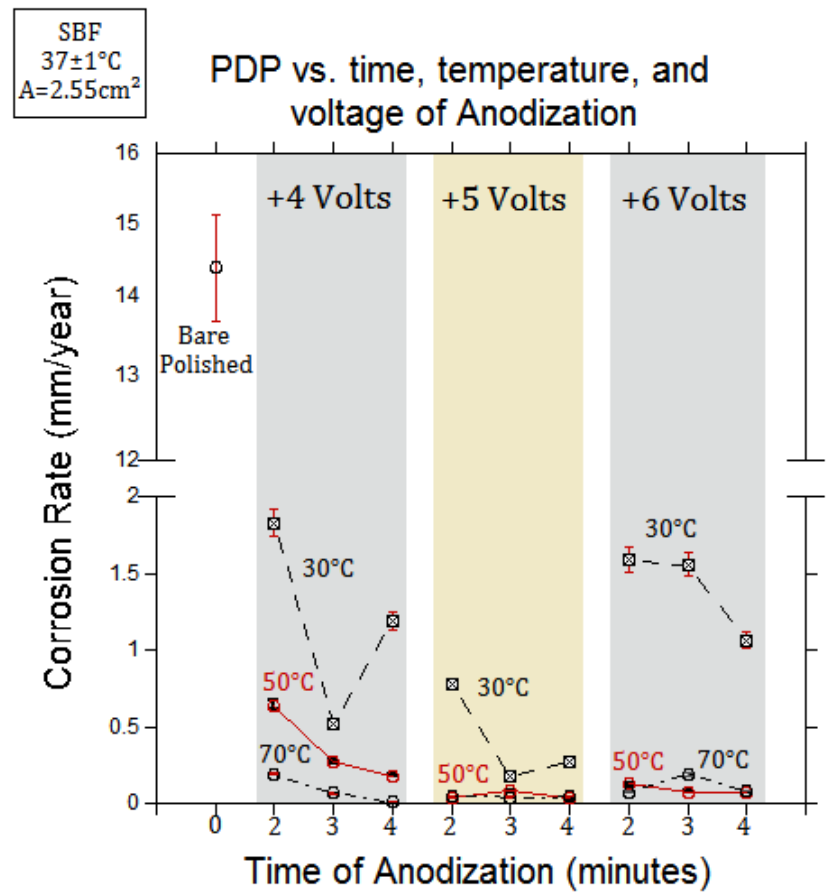

Figure 3. The corrosion rate (mm/year) extracted from PDP curves of coated, anodized HP-Mg specimens at different voltages, times, and temperatures of anodization compared to polished bare specimen.

The anodic polarization curves reveal that the corrosion rate of HP-Mg was significantly decreased in SBF by the formation of the hydrophobic layer of SA on the magnesium substrate. The high corrosion rate of bare HP-Mg (~15 mm/year) was expected due to high concentration of chloride ions (148.8 mM) in SBF. The hydrophobic coating on the anodized layer improved the corrosion resistance of HP-Mg. In general, the PDP results illustrated that the further improvement was achieved by increasing the temperature of the borate benzoate electrolyte during anodization. This might be due to the formation of the high porosity of the hydroxide/oxide layer, and consequently improving the binding between the magnesium substrate and the SA [23].

---
[3] Polished-Bare

The results in Figure 3 also show that the corrosion resistance was improved by increasing the time and voltage of anodization. This might be due to the formation of a thicker hydroxide/oxide layer. However, the optimum corrosion rate (12 $\mu$m/year) belongs to the anodized HP-Mg with +4 volts for 4 minutes at 70°C. The EIS test was performed before the PDP test in SBF at 37°C. The results of EIS test are illustrated in Figure 4 as a function of different anodization parameters (voltage, time, and temperature of anodization) for the anodized coated HP-Mg specimens compared to the polished bare magnesium. Figure 4 (a-c) shows the Bode plot of EIS measurements at different voltages of anodization, +4, +5, and +6 volts, respectively. The coated HP-Mg, anodized with +4 volts for 4 minutes at 70°C, provided the highest polarization resistance ($R_p \simeq 10^5 \Omega$) among all coated specimens (Figure 4-a). The summary of the corrosion results from the EIS and PDP measurements are given in Table 4.

Table 4. Summary of the corrosion results from EIS and PDP meeasurments (A=2.55 cm²).

| time (min) | T (°C) | +4 Volts PDP Test $E_{corr}$ (V) | $i_{corr}$ ($\mu$A) | EIS Test $R_p$ (k$\Omega$) | +5 Volts PDP Test $E_{corr}$ (V) | $i_{corr}$ ($\mu$A) | EIS Test $R_p$ (k$\Omega$) | +6 Volts PDP Test $E_{corr}$ (V) | $i_{corr}$ ($\mu$A) | EIS Test $R_p$ (k$\Omega$) |
|---|---|---|---|---|---|---|---|---|---|---|
| 2 | 30 | -1.85 | 277 | 0.16 | -1.79 | 165 | 0.21 | -1.79 | 162 | 0.38 |
|   | 50 | -1.84 | 81 | 0.43 | -1.60 | 2.9 | 16.51 | -1.63 | 19.1 | 3.57 |
|   | 70 | -1.77 | 19.3 | 2.9 | -1.59 | 5.6 | 6.75 | -1.61 | 7.2 | 12.42 |
| 3 | 30 | -1.75 | 35.9 | 1.38 | -1.77 | 25.2 | 4.54 | -1.80 | 168 | 0.53 |
|   | 50 | -1.77 | 23 | 2.9 | -1.67 | 3.1 | 11.12 | -1.63 | 8.1 | 8.64 |
|   | 70 | -1.68 | 3.9 | 17.8 | -1.62 | 3.5 | 16.49 | -1.58 | 17.2 | 3.69 |
| 4 | 30 | -1.52 | 100 | 0.24 | -1.83 | 57.9 | 2.41 | -1.79 | 242 | 1.16 |
|   | 50 | -1.72 | 26.1 | 2.72 | -1.57 | 1.7 | 19.98 | -1.60 | 5.2 | 10.76 |
|   | 70 | -1.65 | 0.7 | 107.86 | -1.58 | 7.5 | 7.79 | -1.64 | 6.9 | 10.55 |

In order to find the best electrolyte to anodize magnesium, different types of electrolytes (KOH, NaOH, and NaBz) were investigated in this work. The polished (1200 grit) HP-Mg specimens were anodized with +4 volts for 2 minutes at different temperatures (30, 50, and 70°C, then coated with SA (0.05 mol/L). The corrosion rates (mm/year) extracted from PDP measurement are shown in Figure 5. The PDP tests were performed after 15 minutes immersion in warm (37°C) SBF. The results show that the corrosion resistance of anodized HP-Mg specimens in KOH and NaBz were improved by increasing the temperature. However, the NaOH showed the opposite effect in that the corrosion rate increased with increasing the temperature of the electrolyte.

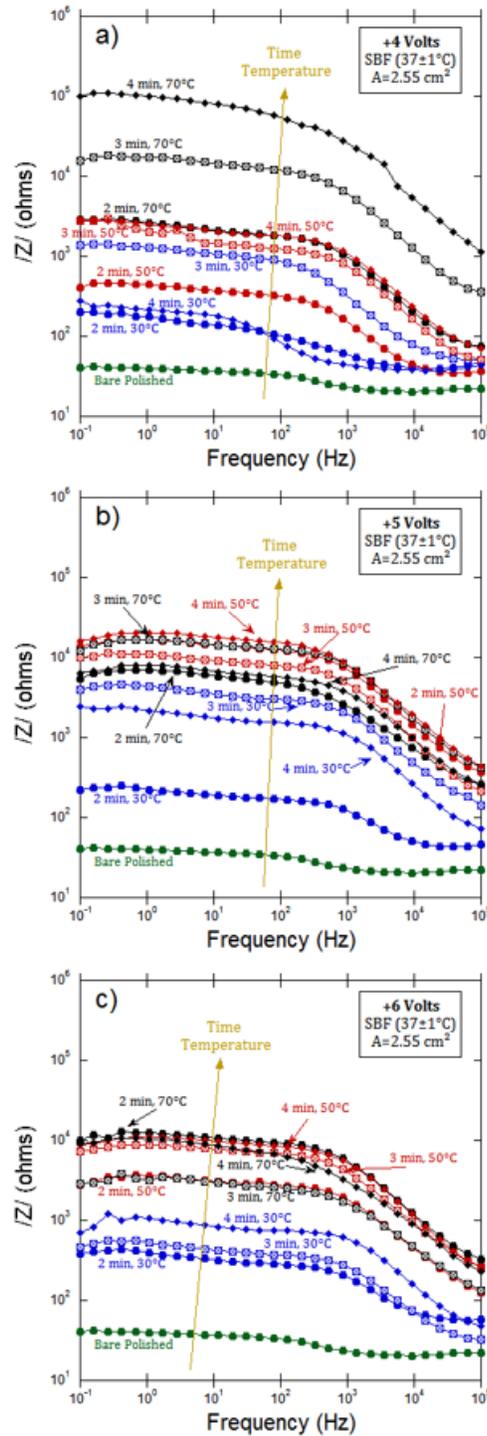

Figure 4. Bode plot of EIS measurement at different voltages of anodization as a function of time and temperature of anodization process; a) +4 Volts, b) +5 Volts, and c) +6 Volts.

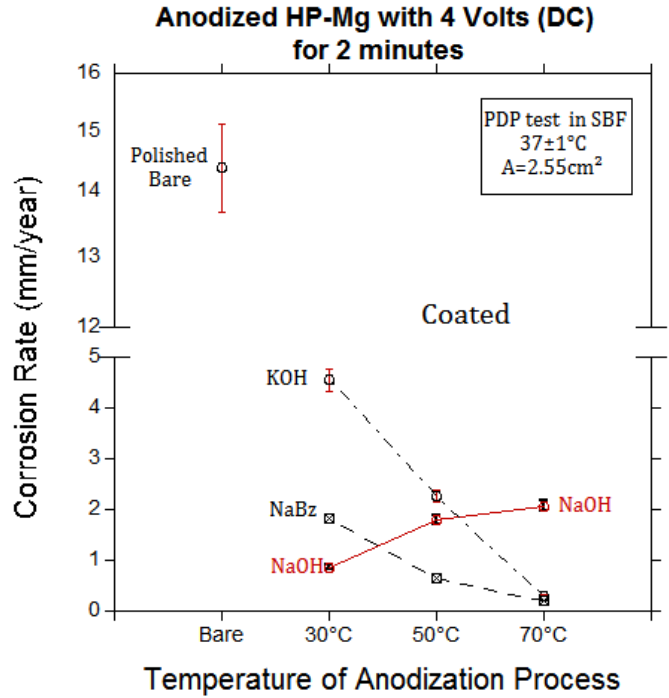

*Figure 5. The corrosion rate (mm/year) extracted from PDP measurements of coated HP-specimens as a function of the electrolyte and temperature.*

The PDP curves in Figure 6 did not depict a very different anodic polarization behavior for the coated samples which were immersed in the different SA concentrations. The coated specimens were polished (1200 grit), anodized with +4 volts for 2 minutes in warm borate benzoate (30°C), and then immersed in 0.05, 0.1, and 0.15 mol/L SA in ethanol for 60 minutes. The results (Figure 6) reveal that the increasing SA concentration increased the corrosion potential but had a negligible effect on the corrosion resistance. This might be because the SA reduced the anodic sites but did not reduce the cathodic sites. Therefore, the excess concentration of SA was not more effective at enhancing the corrosion resistance of magnesium.

The effect of surface preparation on the corrosion behavior of the anodized HP-Mg specimens was also investigated. HP-Mg specimens were polished to 1 $\mu$m in order to compare with the polished specimens to 1200 grits. Then anodized with +4 volts for 2 and 3 minutes in borate benzoate at 30°C, immersed in 0.05 mol/L SA solution. The corrosion rates from PDP measurements suggest that the 1 $\mu$m diamond polish improved the corrosion resistance of HP-Mg in SBF at 37°C (see Figure 7). The results show that the corrosion resistance of 1 $\mu$m polished magnesium was enhanced 10x and 5x over the anodized specimens with +4 volts for 2 and 3 minutes respectively. The smoother surface can improve the efficiency of the anodizing process.

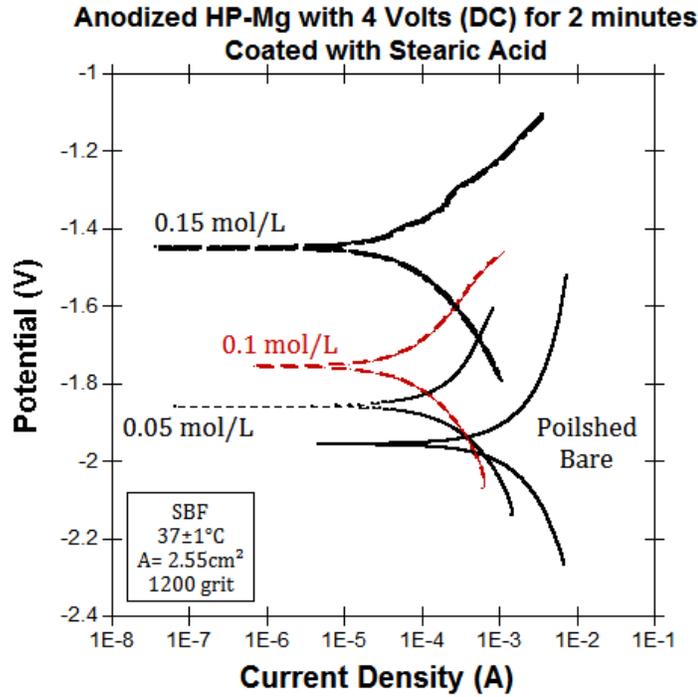

*Figure 6. The potentiodynamic polarization curves in SBF at 37°C for different concentrations of SA.*

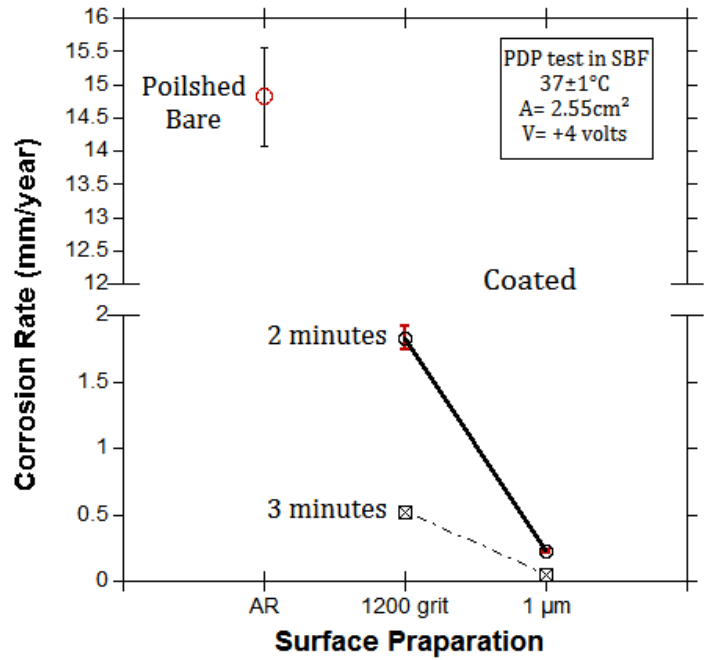

*Figure 7. The corrosion rate (mm/year) extracted with PDP measurements as a function of surface preparation and time of anodization at +4 volts.*

### 3.3. Long-term Corrosion Behavior

In order to investigate the coated layer performance in SBF with the passing of time, a coated HP-Mg specimen ( +4 volts, 2 minutes borate benzoate at 30°C) was immersed in SBF at 37°C for 72 hours, and EIS was conducted at different times of immersion (t=0, 10 minutes, 1, 2, 4, 12, 24, 48, and 72 hours). The PDP test alters the surface, therefore, the EIS test is usually employed for investigating the long-term corrosion behavior of immersed specimens [20]. Figure 8 shows that the coated HP-Mg lost its protectiveness after 12 hours immersion in SBF, and then regained some of it.

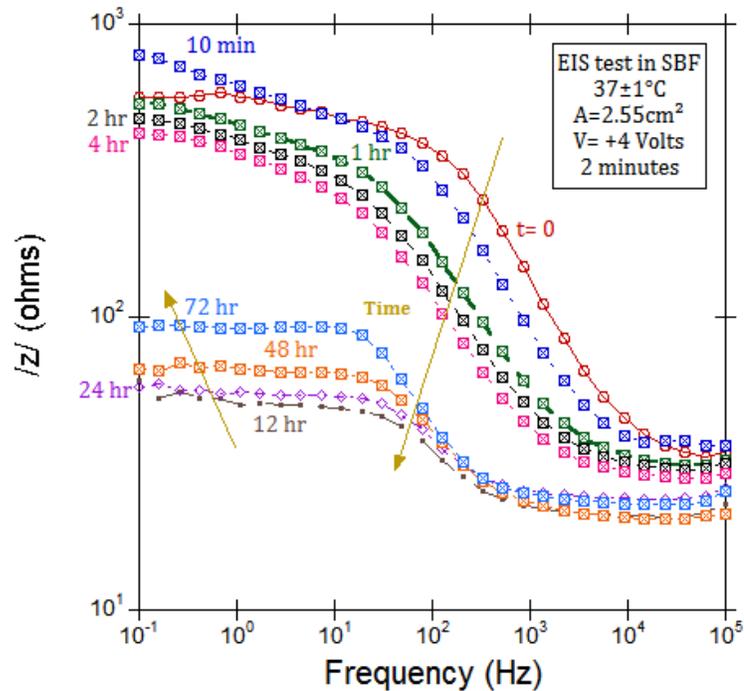

*Figure 8. The EIS results for the sample was anodized in borate benzoate (30°C, +4 volts, 2 minutes) after immersion in SBF for 10 minutes, 1 hr., 2 hr., 4 hr., 12 hr., 24 hr., 48 hr., and 72 hr. at 37°C.*

## 4. Conclusions

In order to improve the corrosion resistance of biodegradable magnesium implants, the magnesium was anodized and then coated with SA. The effect of anodization parameters such as voltage, time, temperature, and electrolyte of anodization process, and also surface smoothness and concentration of SA solution were investigated in this research.

The corrosion behavior of coated and bare HP-Mg were studied by EIS and PDP in SBF electrolyte at 37°C. We demonstrated the following: The SA layer on an HP-Mg substrate provided super hydrophobicity with 164° water contact angle. The coated layer of SA significantly improved the corrosion resistance of HP-Mg. Increasing the voltage and time of anodization further improve the corrosion resistance. The corrosion rate of HP-Mg was dramatically decreased by increasing the anodization temperature to 70°C. The optimum result was anodizing 1200 grit polished HP-Mg at +4 volts for 4 minutes in hot borate benzoate (70°C), and the corrosion rate was significantly decreased to 12 $\mu$m/year in SBF. Polished bare HP-Mg had a corrosion rate 1000x higher at 15 mm/year.

The borate benzoate was the best choice to anodize magnesium at +4 volts for 2 minutes at 50 and 70°C compared to KOH and NaOH. The change in concentration of SA solution did not significantly affect the corrosion behavior of coated HP-Mg.

The 1 $\mu$m polished specimens provided 10x and 5x higher corrosion resistance with +4 volts anodized HP-Mg for 2 and 3 minutes respectively compared to the 1200 grit polished specimens. The long-term results illustrated that coating layer was a temporary layer and its protectiveness decreased over time.